\documentclass[journal]{IEEEtran}

\usepackage{graphicx}
\usepackage{amsmath}
\usepackage{amssymb}
\usepackage{xcolor}
\usepackage{verbatim}
\usepackage{hyperref}
\usepackage{algorithm}
\usepackage{algpseudocode}
\usepackage{listings}
\usepackage{caption}
\usepackage{subcaption}
\usepackage{authblk}

\usepackage[noadjust]{cite}
\usepackage[labelformat=simple]{subcaption}
\begin{document}

\title{CTformer: Convolution-free Token2Token Dilated Vision Transformer for Low-dose CT Denoising}

\author{Dayang Wang$^1$, Fenglei Fan$^2$, Zhan Wu$^1$, Rui Liu$^3$, Fei Wang$^2$, Hengyong Yu$^{1*}$
% ~\IEEEmembership{Senior Member,~IEEE}
\thanks{$*$ Dr. Hengyong Yu is the corresponding author.}
}
\affil{
$^1$Department of Electrical and Computer Engineering, University of Massachusetts, Lowell, MA, USA\\
$^2$ Weill Cornell Medicine, Cornell University, New York City, NY, US\\
$^3$ 3920 Mystic Valley Parkway, Medford, MA, US
% $^3$ Mathworks, Inc.
% $^3$ 3920 Mystic Valley Parkway, Medford, MA, USA
}

% \markboth{Please submit the manuscript to the Special Issue on Explainable and Generalizable Deep Learning for Medical Imaging.}%
% {Shell \MakeLowercase{\textit{et al.}}: Bare Demo of IEEEtran.cls for IEEE Journals}

\maketitle

\begin{abstract}
Low-dose computed tomography (LDCT) denoising is an important problem in CT research. Compared to the normal dose CT (NDCT), LDCT images are subjected to severe noise and artifacts. Recently in many studies, vision transformers have shown superior feature representation ability over convolutional neural networks (CNNs). However, unlike CNNs, the potential of vision transformers in LDCT denoising was little explored so far. To fill this gap, we propose a Convolution-free Token2Token Dilated Vision Transformer (CTformer\footnote{This manuscript is an extension of our conference paper \cite{10.1007/978-3-030-87589-3_43}. }) for low-dose CT denoising. The CTformer uses a more powerful token rearrangement to encompass local contextual information and thus avoids convolution. It also dilates and shifts feature maps to capture longer-range interaction. We interpret the CTformer by statically inspecting patterns of its internal attention maps and dynamically tracing the hierarchical attention flow with an explanatory graph. Furthermore, an overlapped inference mechanism is introduced to effectively eliminate the boundary artifacts that are common for encoder-decoder-based denoising models. Experimental results\footnote{Codes are available at 
\url{github.com/wdayang/CTformer}} on Mayo
LDCT dataset suggest that the CTformer
outperforms the state-of-the-art denoising methods with a low computation overhead.
\end{abstract}
% Codes are available at https://github.com/wdayang/TED-net

% \fi
\IEEEpeerreviewmaketitle

\section{INTRODUCTION}

The LDCT problem has gained lots of attention in the community due to its potential of reducing X-ray radiation. However, compared to NDCT images, LDCT images suffer from severe noise and artifacts \cite{RN98} when they are applied to clinical applications. To overcome this problem, two types of algorithms have been investigated: traditional algorithms and convolutional neural networks (CNNs) \cite{he2016deep,fan2021sparse}. i) Traditional algorithms such as iterative methods suppress the artifacts and noise by using a physical model based on a certain prior. Unfortunately, these algorithms are hard to be adopted in commercial CT scanners because of the hardware limitations and high computational cost \cite{RN54}. ii) With the advent of deep learning, CNNs have been a prevailing approach for LDCT image denoising. Despite the superior learning ability aided by big data\cite{RN100}, CNNs are reported to be limited in capturing long-range contextual information in images \cite{wang2018non,RN57,RN74,RN76}, which will adversely affect the retrieval of richer structural information in denoised images. 

\begin{figure*}
\centering
\includegraphics[width=\textwidth]{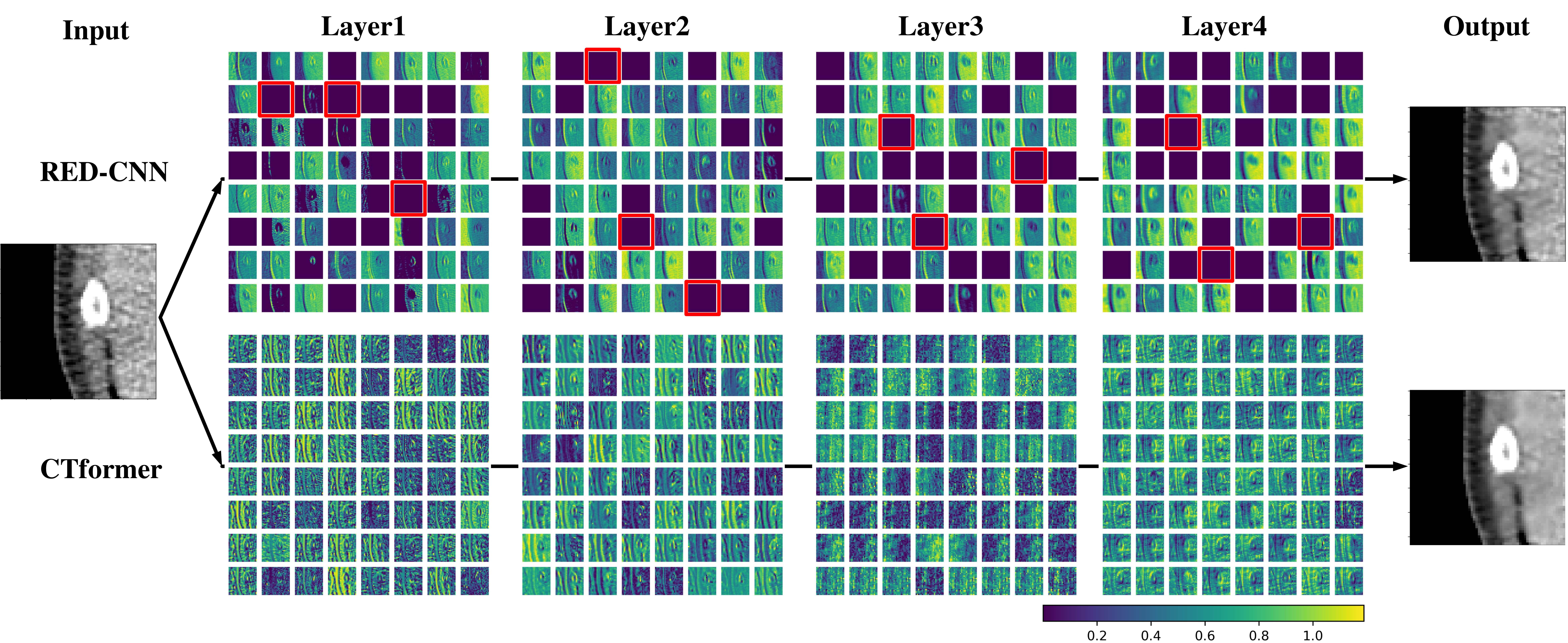}  %% intermediate_map_v3, intermediate_map_v5 (gray feature map)
\caption{The feature maps visualization of the pretrained RED-CNN and the CTformer. The transformer model (CTformer) has diversified and effective features, while the CNN model (RED-CNN) has a lot of inactive features.}
\label{featuremap}
\end{figure*}

Recently, the transformer model \cite{RN57} has shown excellent performance in computer vision \cite{RN66,RN63,RN65,RN56,RN79,RN70,RN73,RN75,RN61,RN78,choromanski2020rethinking}. Dosovitskiy \textit{et al.} proposed the first vision transformer (ViT) by simply mapping an image into 16$\times$16 patches (this operation is commonly referred to as tokenization) in analogy to words in a sentence in natural language processing \cite{RN56}. Yuan \textit{et al.} further proposed a Token2Token method to empower the transformer model with a diverse information encoding \cite{RN76}. Next, Liu \textit{et al.} designed a swin transformer to include patch fusion and cyclic shift to enlarge the perception of contextual information in tokens \cite{RN74}. Moreover, Choromanski \textit{et al.} proposed a Performer transformer to reduce the computational complexity of the self-attention by approximating the inherent softmax operator \cite{choromanski2020rethinking}.
% Other than these methodological improvements, researchers also applied the transformer to low-level vision tasks \cite{RN65,RN79,RN70}. 
Currently, the transformer model is poised to replace CNNs as the mainstream deep learning model. On the one hand, compared to CNNs, the transformer model is good at capturing global information and long-range feature interactions, resulting in the utilization of richer information. As shown in Fig. \ref{featuremap}, the transformer has diversified and effective features, while the CNN model has many inactive features. On the other hand, the transformer model enjoys higher visual interpretability by the virtue of its inherent self-attention block \cite{abnar2020quantifying,montavon2019layer,chefer2021transformer}. However, a typical CNN model contains no generic explanation modules \cite{fan2021interpretability}.

Despite the success and great promise, the transformer has been little investigated in LDCT denoising. In our opinion, the transformer model is suitable for LDCT denoising problem. Other than the effectiveness, a transformer is more desirable for physicians because it is self-explanatory \cite{singh2020explainable}, \textit{e.g.}, allowing a physician to make sense of the model's logic. To the best of our knowledge, Zhang \textit{et al.} pioneered to apply the transformer in LDCT denoising \cite{RN77}. Although this model achieves the state-of-the-art performance, it has imperfections in three aspects: i) The model uses the vanilla transformer which can not fully explore the potential of the transformer, as relevant studies are rapidly advancing. ii) Intensive convolutions are included in the model, making their model essentially a hybrid model. Thus, the merits of using a transformer are insufficiently justified. iii) Their work neglects the interpretability that is essential for clinical applications \cite{kim2019visual}.

We aim to fully explore the potential of transformers in LDCT denoising. Specifically, we propose a Convolution-free Token2Token Dilated Vision Transformer (CTformer) for low-dose CT denoising. The CTformer has the following characteristics: i) Although the convolution is instrumental to capture local features when it is combined with transformers on small datasets, it is not a necessity for the performance because the token rearrangement can also help complement the local information. Therefore, we completely exclude convolution operations in the proposed CTformer. To the best of our knowledge, the CTformer is the first pure transformer for LDCT denoising. ii) The dilation and a cyclic shift are used in the Token2Token to enlarge the receptive field, thereby gaining broader contextual information from the feature maps and reducing the computational cost. iii) We utilize an overlapped inference mechanism to address the boundary artifact that is common in the encoder-decoder denoising models. iv) We develop interpretability for the CTformer with the visual attention maps and an explanatory graph that shed light on how the CTformer discriminates key structures from noise as well as hierarchical attention flow across layers.
Experiments results suggest that the CTformer delivers superior denoising performance over other state-of-the-arts with fewer trainable parameters and multiply-accumulate operations (MACs). 

In summary, our contributions are threefold: i) This work is the forerunner to apply the vision transformer to LDCT denoising problem. What's more, the proposed CTformer is the first pure transformer. ii) We introduce dilation and cyclic shift to enhance the tokenization process in the model, utilize a new inference mechanism to fix the boundary artifacts, and develop the interpretation methods to unveil the model's denoising patterns.
iii) Our experimental results demonstrate the superior denoising performance and model efficiency of the CTformer for LDCT denoising.

% Please note that this manuscript is an extension of our previously-published conference paper \cite{10.1007/978-3-030-87589-3_43}. 
% We elaborate the CTformer with more detailed methodological and experimental analyses. 

\section{RELATED WORK}
The previous studies for the LDCT denoising problem can be categorized into two classes.

\textbf{Traditional algorithms.} Typically, these methods incorporate a physical prior into an iterative reconstruction framework to suppress noise. 
% traditional methods, e.g., iterative methods, propose to suppress the artifact and noise by using a physical model that is based on a certain prior. 
For example, compressed sensing (CS) has been widely used for the LDCT problem by adopting a sparse representation \cite{RN43}, \textit{i.e.}, the total variation minimization assumes that the clean image is piecewise constant whose gradients are sparse \cite{RN44,RN45,RN46,RN47}. Xu \textit{et al.} used a dictionary to construct the sparse representation \cite{RN48} for LDCT denoising. In addition to the sparsity prior, Ma \textit{et al.} designed a non-local mean prior to utilize the image voxels across the whole image rather than the local region \cite{RN53}. However, increasingly more studies \cite{wu2017iterative,RN69,RN68,RN72} implied that the traditional algorithms are surpassed by deep learning models driven by big data. 
% Nonetheless, these algorithms are hard to be adopted in commercial CT scanners because of the hardware limitations and high computational cost \cite{RN54}.

%%(\textit{e.g.} ResNet, DenseNet, S3Net) 
\textbf{Convolution models.}
CNNs have been used for the LDCT image reconstruction. Wu \textit{et al.} used a K-sparse autoencoder to learn the image features in an unsupervised fashion and minimize the distance between a normal-dose image and an iterative reconstruction result in the feature space of the autoencoder \cite{wu2017iterative}. Liu \textit{et al.} proposed a 3D residual convolutional network to estimate an iterative reconstruction (IR) image from an LDCT analytic reconstruction image \cite{liu2019deep}. Their method can save time because it avoids the time-consuming iterative reconstruction. He \textit{et al.} proposed the 3pADMM method to address the problems of hyper-parameter optimization and prior knowledge selection in LDCT reconstruction \cite{RN95}. 

Besides, a majority of deep LDCT denoising models focused on image post-processing. The paper of Chen \textit{et al.} was a pioneer work which employed the convolution, deconvolution, and shortcut connections to prototype a residual encoder-decoder convolution neural network (RED-CNN) \cite{RN69}. Yang \textit{et al.} used the generative adversarial network with Wasserstein distance (WGAN) aided by a perceptual loss to improve the quality of denoised images \cite{RN68}. Due to the excellent performance of WGAN in generating faithful real-world CT images and the role of the perceptual loss in structural fidelity, this model alleviated the over-smoothness in the denoised images. Li \textit{et al}. employed a GAN armed with the structural similarity loss, the perceptual loss, the adversarial loss, and the sharpness loss to preserve structural details and sharp boundaries \cite{li2021low}. Fan \textit{et al.} constructed a quadratic neuron-based autoencoder for LDCT image denoising with more robustness and efficiency as opposed to conventional CNN-based methods \cite{RN72}. It is the first autoencoder based on a new type of neurons. Huang \textit{et al}. proposed a two-stage residual CNN \cite{huang2020two}, where the first stage uses stationary wavelet transform for texture denoising, and the second one enhances the image structure via combining the average of NDCT images and the denoised image from the first stage.

However, CNN-based models typically lack the ability to capture global contextual information due to the limited receptive fields, thus less efficient to model the structural similarity across the whole image \cite{RN77,10.1007/978-3-030-87589-3_43,wang2021dudotrans}.

\section{METHODS}

\begin{figure}
\centering
\includegraphics[width=\linewidth]{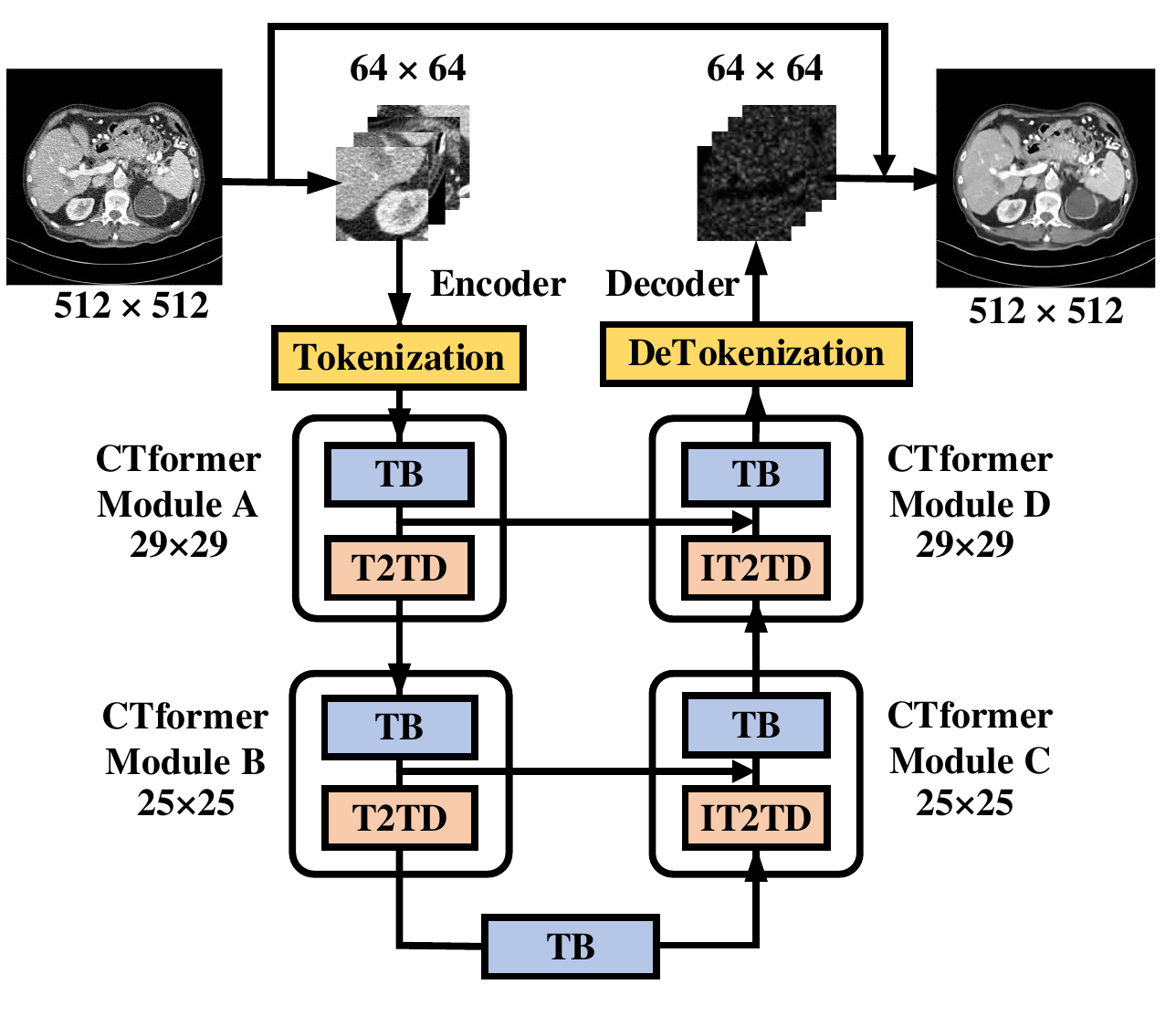}
\caption{The CTformer consists of the residual encoder-decoder structure with  tokenization/detokenization blocks, four CTformer modules with different sizes of feature maps, and an intermediate transformer block. 
Tokenization block unfolds image patches into sequential tokens, while detokenization block converts tokens back to the image. Each encoder CTformer module includes  a transformer block (TB) and a Token2Token dilation block (T2TD), while each decoder CTformer module consists of an inverse Token2Token dilation block (IT2TD) and a TB, symmetrically. 
% CSB uses cyclic shift operation and ICSB employs inverse cyclic shift. T2T incorporates Token2Token block to enhance tokenization while T2TD includes dilation in the T2T tokenization process.
} \label{fig:whole}
\end{figure}

In the supervised setting, with a deep learning model, the LDCT denoising task is to learn a mapping from a paired noisy LDCT image $\boldsymbol{x}$ to a clean NDCT image $\boldsymbol{y}$. 
Mathematically, a neural network can be trained by optimizing a mean square error (MSE) loss function as follows:

\begin{equation}
   \underset{\boldsymbol{W}}{ \min} \ \  \mathcal{J}(\boldsymbol{W};\boldsymbol{x})=\left \| f(\boldsymbol{W};\boldsymbol{x})-\boldsymbol{y} \right \|_2,
    \label{equ1}
\end{equation}
where $f(\boldsymbol{W};\boldsymbol{x})$ is a neural network, and $\boldsymbol{W}$ is a collection of parameters for simplicity. 

\subsection{Architecture of the CTformer}
As shown in Fig. \ref{fig:whole}, the proposed CTformer takes the residual encoder-decoder structure with tokenization/detokenization blocks, four CTformer modules, and an intermediate transformer block. In the encoder, CTformer modules A and B include a transformer block (TB), and a Token2Token Dilation block (T2TD). In the decoder, CTformer modules C and D symmetrically encompass an inverse Token2Token Dilation block (IT2TD) and a TB. The IT2TB block takes the inverse design of the corresponding T2TB block. Now let us introduce the CTformer from its macro to micro structures.

\begin{figure*}%[a]{0.4\textwidth}
     \centering
     \includegraphics[width=\linewidth]{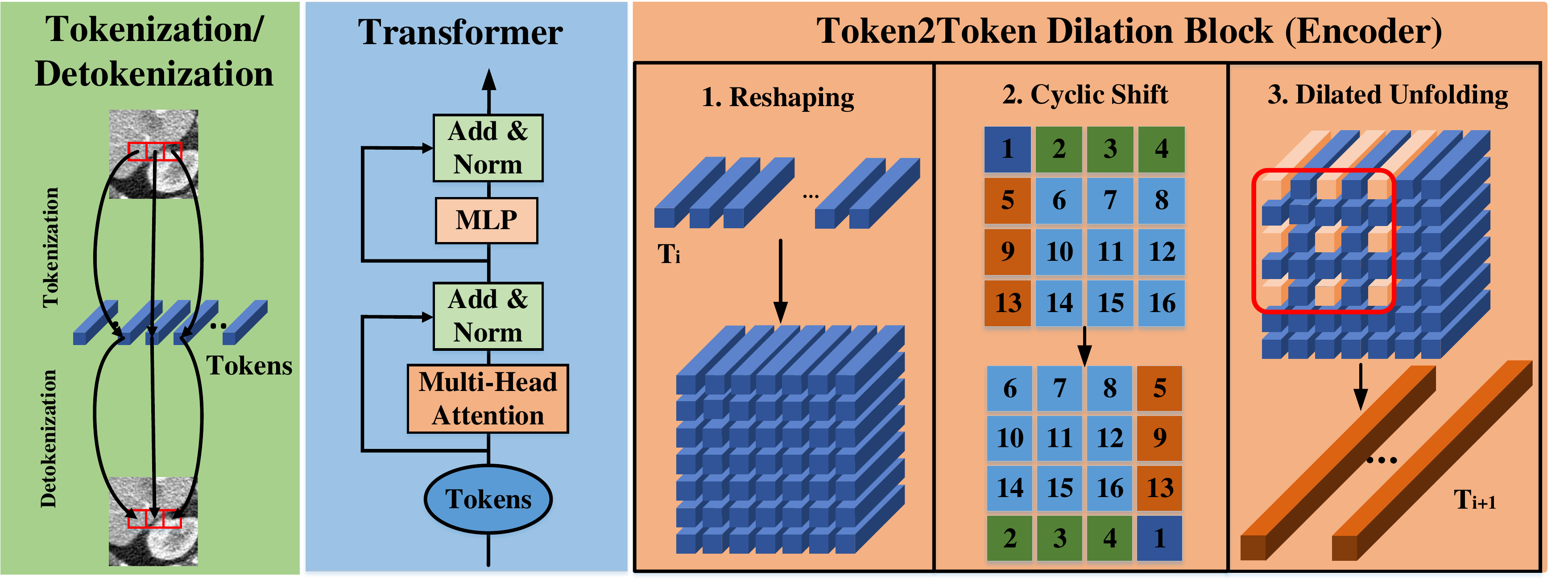}
     \caption{The micro structures of the CTformer: tokenization/detokenization, transformer block and Token2Token dialtion block.}
     \label{micro}
\end{figure*}

\textbf{Residual encoder-decoder structure.} We use a residual encoder-decoder structure as the backbone of the CTformer. The shortcuts only bridge similar levels of layers in encoder and decoder parts. Although the unsatisfactory information loss is accompanied by denoising in the encoder block, which hurts the structural recovery in the decoder part, the employment of shortcuts can supplement information from the feature maps of the encoder to retain structural details. Besides, shortcuts can fix the gradient vanishing problem such that a deep model can still be stably trained \cite{he2016identity}.

\textbf{Tokenization block.}
As shown in Fig. \ref{micro}, in the tokenization process, a noisy CT image is unfolded into a sequence of two dimensional (2D) patches (also referred to as tokens): $\mathbf{T_0} \in \mathbb{R}^{b \times n \times d_0} $, where $b$ is the batch size, $n$ is the number of tokens, and $d_0$ is the token dimension. Throughout this manuscript, we use tokens and patches interchangeably.

\textbf{Transformer block.}
As shown in Fig. \ref{micro}, 
% a multiple layer perceptron (MLP) extends the token embedding to the desired size before transmitted to the latter transformer, which 
a typical transformer block contains multiple head attention (MHA), layer normalization (LN), an MLP, and residual connections to enhance the expressive power. Specifically, in the self-attention, a token sequence $\mathbf{T_0} \in \mathbb{R}^{b \times n \times d_0}$ is linearly mapped into three tensors which are respectively referred to as query, key, and value, denoted as $\mathbf{Q}, \mathbf{K}, \mathbf{V} \in \mathbb{R}^{b \times n \times d_m}$ for short, where $d_m$ is the token embedding dimension. Mathematically, we have 
\begin{equation}
\begin{cases}
    & \mathbf{Q} =\mathbf{T_0} \mathbf{W_q}   \\
    & \mathbf{K} =\mathbf{T_0} \mathbf{W_k}   \\
    & \mathbf{V} =\mathbf{T_0} \mathbf{W_v} , \\
\end{cases}
\label{eqn:attention_block}
\end{equation}
where $\mathbf{W_q}$, $\mathbf{W_k}$ and $\mathbf{W_v}$ are linear operators. 
Then, the output of the self-attention is calculated as
\begin{equation}
    \mathrm{MHA}(\mathbf{Q},\mathbf{K},\mathbf{V}) = \mathrm{softmax}(\frac{\mathbf{Q}{{\mathbf{K}}^\top}}{\sqrt{{{d}_{k}}}}) \mathbf{V},
\label{MHA}
\end{equation}
where the scaling factor $\frac{1}{\sqrt{{{d}_{k}}}}$ is based on the network depth. 
% Eq. \eqref{MHA} has several variants, e.g., 
% \begin{equation}
%     \widehat{\mathrm{MHA}} = \mathbf{Q}' (\mathbf{K}'^{\top} \mathbf{V}),
% \end{equation}
% where $\mathbf{Q}' \in \mathbb{R}^{b \times n \times d_p}$ and $\mathbf{K}' \in \mathbb{R}^{b \times n \times d_p}$ are approximated from $\mathbf{Q}' \mathbf{K}' \approx \mathrm{softmax}(\frac{\mathbf{Q}{{\mathbf{K}}^\top}}{\sqrt{{{d}_{k}}}})$ by a kernel method. Thus, the computational cost of Eq. \eqref{MHA} is reduced. The transformer using this approximation is also called Performer \cite{choromanski2020rethinking}.
Besides the authentic calculation of Eq. \eqref{MHA}, 
% which requires quadratic complexity to $n$, 
the $\mathrm{softmax}$ operator can be approximated by a kernel method, thus, obtaining a reduced complexity of Eq. \eqref{MHA}. The transformer using this approximation is also called Performer \cite{choromanski2020rethinking}.

% Eq. \eqref{MHA} has several variants, e.g., 
% \begin{equation}
%     \mathrm{MHA}(\mathbf{Q},\mathbf{K},\mathbf{V}) = \mathbf{D}^{-1} \mathbf{A} \mathbf{V},
% \label{performer}
% \end{equation}
% where $\mathbf{A} = \mathrm{exp}(\mathbf{QK^\top}/\sqrt{{d}_{k}})$, $\mathbf{D} = \mathrm{diag} (\mathbf{A} \mathbf{1}_{n})$,  $\mathrm{diag(\cdot)}$ is the diagonal matrix, and $\mathbf{1}_{\mathrm{n}}$ is the all-ones vector of length $n$. The transformer using Eq. \eqref{performer} is also called Performer \cite{choromanski2020rethinking}.

$\mathrm{\mathbf{Att}}=\mathrm{softmax}(\mathbf{Q} \mathbf{K}^\top / \sqrt{d_k})$ is the attention map that will be used in the post-hoc interpretability analysis. Through the transformer block, the output token $\mathbf{T}_a \in \mathbb{R}^{b \times n \times d_a}$ is 

\begin{equation}
\begin{cases}
   & \mathbf{T}_a^{'}=\mathrm{MHA}(\mathrm{LN}( \mathrm{MLP} (\mathbf{T}_0))) + \mathbf{T}_0 \\
   & \mathbf{T}_a=\mathrm{MLP}(\mathrm{LN}(\mathbf{T}_a^{'})) + \mathbf{T}_a^{'}.
\end{cases}
\end{equation}

% with a scaling factor of $\frac{1}{\sqrt{{{d}_{k}}}}$, and
% \begin{equation}
%     Q = W_q T,
%     K = W_k T,
%     V = W_v T,
% \end{equation}
% where $W_q,W_k$ and $W_v$ are linear operators. 
% When MHA is used, the output is then calculate for $m$ heads:
% \begin{equation}
%     Q_i = W^i_q T, K_i = W^i_k T, V_i = W^i_v T, i = 1,2...,m,
% \end{equation}
% \begin{equation}
%     Head_i = Attention(Q^i,K^i,V^i),
% \end{equation}
% \begin{equation}
%     MHA(T) = \big(Concat_{i=1}^m(Head_i) \big)W^o,
% \end{equation}
% where $W^i_q,W^i_k,W^i_v$ and $W^o$ are linear operators. 

\textbf{Token2Token dilation block.}
Previously, the simple tokenization in the vanilla transformer only includes one tokenization process using either reshaping or convolutions with a fixed stride to convert an image to tokens. Thus, it tends to ignore the dependence across neighboring tokens. What's worse, it also makes the attention expressions redundant, which adversely results in limited feature richness in each layer \cite{RN76}. To overcome these problems, as shown in Fig. \ref{micro}, we adopt the recently-proposed T2T block which uses cascade tokenization to replace the simple tokenization \cite{RN76}. The T2T block consists of reshaping and unfolding which can not only model the local information from the surrounding image pixels but also gain more feature representation than convolution. Furthermore, we use cyclic shift and dilation in the T2T (T2TD) to refine the contextual information fusion and leverage spatial relations across a larger region. Now, let us elaborate on these operations in detail.

\underline{Step 1: reshaping}. A sequence of tokens $\mathbf{T_a} \in \mathbb{R}^{b\times n \times d_a}$
given rise by the transformer block are first transposed to $\mathbf{T_a}^\top \in \mathbb{R}^{b\times d_a \times n}$ and then reshaped into $\mathbf{F}\in \mathbb{R}^{b\times d_a \times h \times w}$:
\begin{equation}
    \mathbf{F} = \mathrm{reshape}(\mathbf{T_a}^\top),
    \label{reshape}
\end{equation}
where $h=w=\sqrt{n}$ are the height and width of the feature map, respectively.

\underline{Step 2: cyclic shift}.
We employ the cyclic shift to modify the 4D feature maps in each T2TD block. Specifically, the pixel values in the feature maps are shifted in a cyclic way to utilize the information more sufficiently. Then, an inverse cyclic shift is performed in the symmetric IT2TD block in the decoder to avoid any pixel mismatch in the final denoising results. Through cyclic shift, the tokens fed into the consequent transformer blocks are extracted from different feature maps rather than the fixed patches. 
Furthermore, now the tokens from the boundaries of the modified feature maps include pixels that are not boundaries in the original feature maps. 
In practice, the CTformer shifts the image by two pixels to extract new tokens. Fig. \ref{micro} 
% \begin{figure}%[a]{0.4\textwidth}
%      \centering
%      \includegraphics[width=\linewidth]{figure/Cyclic_shift3.pdf}
%      \caption{(a) An example of one step cyclic shift on $4\times4$ feature map. (b) The patch area changes across different layers.}
%      \label{fig:cyclic}
% \end{figure}
illustrates the cyclic shift module,\\
\begin{equation}
    \mathbf{F}_{c} = \mathrm{cyclicshift}(\mathbf{F}).
\end{equation}

\underline{ Step 3: dilated unfolding}. The dilated unfolding will use the unfolding operation to retokenize the feature maps from the last step. To alleviate the information loss in this step, we adopt an overlapped splitting of patches. As a result, these aggregated tokens can respect the correlations among the neighboring tokens. 
\begin{equation}
    \mathbf{T}_s = \mathrm{dilatedunfold}(\mathbf{F_c}).
\end{equation}
In this stage, the 4D feature maps $\mathbf{F}\in \mathbb{R}^{b\times d \times h \times w}$  are converted back to 3D tokens $\mathbf{T}_s \in \mathbb{R}^{b\times n_s \times d_s}$, where $n_s$ and $d_s$ represent the new token number and token dimension, respectively. By aggregating surrounding patches and pixels, the local information is favorably preserved, and the number of tokens is changed. Specifically, the token number decreases in the encoder and increases in the decoder.

\begin{figure}
\centering
\includegraphics[width=\linewidth]{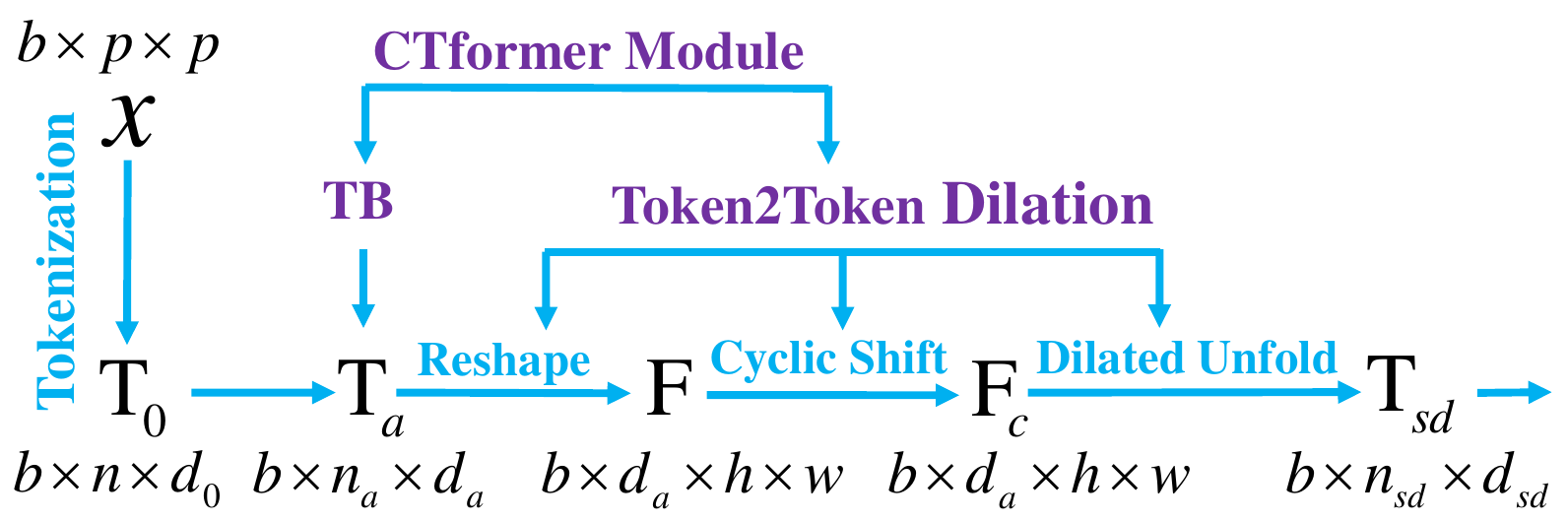}
\caption{A flowchart to illustrate the computation of tensors in the CTformer architecture for readers' reproducibility.} \label{tensor}
\end{figure}

Instead of the normal unfolding, we endow the unfolding with a dilation to capture the longer range contextual information with less computational cost.
% . More favorably, unfolding with dilation not only increases the receptive field but also requires 
% less computational costs. 
Mathematically, the perceptive field $P$ of the dilation can be calculated as follows:
\begin{equation}
    P = \prod_{i=0}^1 (2^{K_i+D_i}-1),
\end{equation}
where $K_i$ and $D_i$ denote the kernel size and the dilation rate in a certain dimension, respectively. After the dilated unfolding, the input feature map $\mathbf{F}\in \mathbb{R}^{b\times d \times h \times w}$ becomes $\mathbf{T}_{sd}\in \mathbb{R}^{b\times n_{sd} \times d_{sd}}$,  where $d_{sd}=d \times \prod_i {K_i}$ and the total number of tokens $n_{sd}$  after the dilated unfolding operation is calculated as:
\begin{equation}
    n_{sd}=  \prod_{i=0}^1{\left \lfloor \frac{\mathrm{spatial}(i)-\mathrm{dilation}\times (K_i-1)-1}{\mathrm{stride}}+1 \right \rfloor},
\label{unfold}
\end{equation}
where $\left \lfloor \cdot \right \rfloor$ is the floor function, $\mathrm{spatial}(i)$ means corresponding size in the $i$ dimension, $\mathrm{spatial}(0) = h$ in height dimension, and $\mathrm{spatial}(1) = w$ in width dimension. Here, $\mathrm{dilation}$, $\mathrm{kernel}$, and $\mathrm{stride}$ are related parameters in the unfolding operation. Then, an MLP is performed to map the embedding dimension to a desired size. 
For better understanding of our model, a flowchart is attached for the above-discussed tensors in Fig. \ref{tensor}.

%  \begin{comment}
% The following Algorithm \ref{alg:euclid} is the sudo code for a CTformer Block,
% \begin{algorithm}
% \caption{The procedure of a CTformer Module which includes a TB, CSB, and T2T Blocks}\label{alg:euclid}
% \begin{algorithmic}[1]
% \State import torch.nn as nn
% \State Given tokens $x \in R^{b,n,d}$. \# b: batch size, n: number of tokens, d: embedding dimension.
% \State $x = MLP(MHA(x))$ \# Transformer Block.
% \State $x = x.reshape(b,c,h,w)$ \# Reshaping process. c: channel, h: height, w: width.
% \State $x = torch.roll(x, shift, dim)$ \# Cyclic shift block: shift is the quantity and dim is the dimension to perform cyclic shift.
% \State $x = nn.Unfold(kernel, stride, dilation)(x)$ \# Soft split process: unfold with certain kernel size, stride size and dilation.
% \State \textbf{return} $x$
% %\EndProcedure
% \end{algorithmic}
% \end{algorithm}
% \end{comment}

% \begin{comment}
% \begin{figure}
% \centering
% \includegraphics[width=0.5\textwidth]{cyclic_v2.pdf}
% \caption{The structures of Cyclic Shift and Inverse Cyclic Shift operations to enrich the tokenization process by fusion different kernel area.} \label{cyclic}
% \end{figure}
% \end{comment}

\subsection{Inference of the CTformer.}
\begin{figure}[h]
\centering
\includegraphics[width=\linewidth]{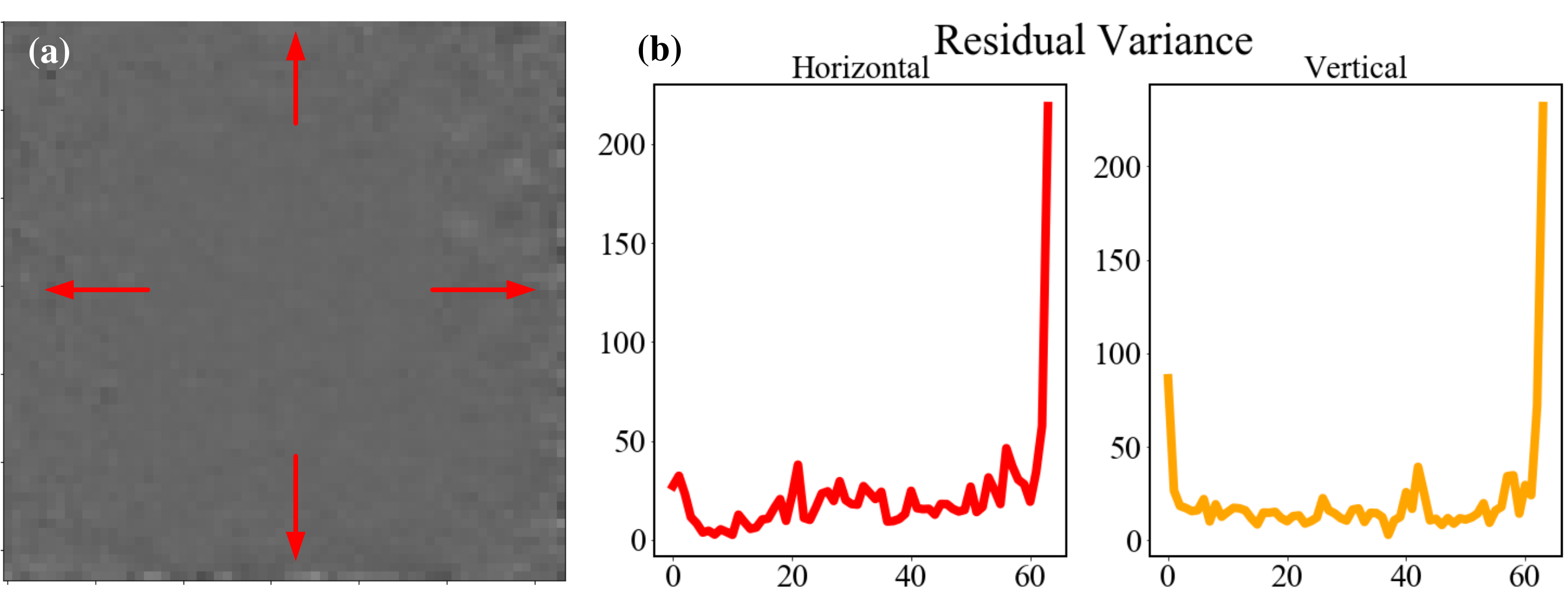}
\caption{(a) The residual map between the prediction and the  NDCT image reveals the boundary artifacts. (b) The profiles of the residual map along the horizontal and vertical axes.
% (a) The denoised results of non-overlapped inference and overlapped inference. (b) The residual map of boundary artifacts for non-overlapped inference. (c) The variance profiles of the residual map on horizontal and vertical axis.
} \label{margin_var}
\end{figure}

In the inference phase, unlike CNN which can directly test the whole image, the transformer model can only do inference patch by patch. Because there exists information loss in the bottleneck of an encoder-decoder architecture \cite{innamorati2020learning}, the denoised results of these patches are inconsistent at boundaries, causing boundary artifacts in the stitched image. As shown in Fig. \ref{margin_var}, we can easily see the mosaic edge indicated by the red arrows, and artifacts are along all four directions. To address this problem, we propose an overlapped inference method. The core of our method is to discard the margin and only keep the center of the model output to stitch the final prediction.

Suppose that the patch size is $p\times p$, we only keep the central part of a patch $(p - 2\eta) \times (p - 2\eta) $ to form the final prediction image, where $\eta$ is selected to be greater than the width of artifacts. In the overlapped inference, slightly more calculations are demanded because we discard the peripheral part of a patch. The increased cost is at the ratio of
\begin{equation}
    \sigma = \Big( \frac{\lceil n/(p-2\eta) \rceil}{\lceil n/p \rceil} \Big )^2, %\times \bigg (\frac{p}{p-2\eta} \bigg )^2,
\label{ratio}
\end{equation}
where $n$ is the original image size, and $\lceil \cdot \rceil$ is the ceiling function. Therefore, we need to balance the computation cost with the artifact elimination effect.

\subsection{Interpretability of the CTformer}
In interpretability research, saliency map is the most popular method. One can generate a saliency map for the CNN-based classification model after the model is trained \cite{selvaraju2017grad}. However, for the image-to-image denoising task, deriving saliency maps are not applicable because denoising models are essentially regression models. In contrast, even if the transformer models are used for denoising, one can leverage the inherent attention modules to achieve saliency maps. Utilizing such an advantage, we develop the interpretability of the CTformer by probing the patterns of the attention maps. Thus, one can decode the inner-working of the CTformer, with an emphasis on the processing of important structural and semantic information. The self-interpretability makes the CTformer uniquely relative to other LDCT denoising models.

% To the best of our knowledge, we conduct extensive experiments to unveil the hidden pattern of the transformer denoising model in Section III. Results show that the attention maps in the CTformer model can discriminate the semantic object parts and the arbitrary noise area in a pretty transparent way. 

Furthermore, we observe that the attention only reflects where the model attends in a static manner, which cannot convey how the attended parts flow across layers in the CTformer. To complement this dynamic information, inspired by  \cite{zhang2021extraction}, we propose to construct an explanatory graph to describe the hierarchical flow of the attention. We take the attended parts as graph nodes and the attention flow as graph edges. Two nodes linked by a edge are usually co-activated and take similar mapping (denoising). Specifically, we first recognize the attended object parts by identifying the peak activations. Then, we build the graph connections between neighboring layers by forwarding a masked feature map and monitoring the high activations. 
% Please note that we do not link nodes of the same layer.

% ($\mathbf{G}$)
% connecting the object part 
% taking an object part as a node and constructing edges by forwarding a masked feature map and monitoring the changes. A edge connects two nodes which are usually co-activated and take the similar mapping (denoising). Please note that we do not link nodes of the same layer.

\textit{Node:} To identify the object part, we provide two pixel-based methods: TopK and local maximum (LM) selection. The TopK extracts the $K$-highest activation across the attention maps, while the LM detects the local maximum activations. 

\textit{Edge:} To construct edges among nodes, we propose to forward a masked feature map. Specifically, given a node (an object part) in a layer, we mask the feature maps and only keep the region around the node. Then, we feed the masked feature maps to obtain the attention map of the next layer. Finally, we extract the highest activation (node) from the obtained attention map and link it to the given node. 
% Finally, we inspect whether the originally-activated regions (nodes) are still activated. If they are activated, we link the given node to them.

By performing the above steps recursively in two subsequent layers, the whole explanatory graph is built to inform us how the attention of the CTformer is shifted.

\section{EXPERIMENTS}
In this part, our model is trained and evaluated on a publicly available dataset. First, we demonstrate the superior denoising performance and the model efficiency of the CTformer over its counterparts. Then, we confirm the effectiveness of the overlapped inference mechanism. Finally, we elaborate on the model interpretability of the CTformer with the aforementioned interpretation methods. 

%We randomly extract 4 image patches of 64×64 from each original image of size 512×512 in every epoch.  
\textbf{Dataset.}
A publicly released dataset from \textit{2016 NIH-AAPM-Mayo Clinic LDCT Grand Challenge}\footnote{https://www.aapm.org/GrandChallenge/LowDoseCT/} \cite{RN83} is used for model training and testing. The dataset includes $2,378$ $3.0$mm slice thickness of low-dose (quarter) and normal-dose (full) CT images from ten anonymous patients. We select the patient L$506$ data for evaluation, while the rest nine patients for model training. Data augmentation is also applied. We generate more training images by randomly rotating ($90$, $180$, or $270$ degrees) and flipping (up/down, left/right) the original image.

\textbf{Experiment settings.} We list the detailed experimental settings in the following:

\begin{itemize}
    \item The experiments are running on Ubuntu 18.04.5 LTS, with Intel(R) Core (TM) i9-9920X CPU @ 3.50GHz using PyTorch 1.5.0 \cite{RN102} and CUDA 10.2.0. 
    The model is trained with four NVIDIA GTX 2080Ti 11G GPUs.
    \item The intermediate transformer block takes the authentic design, while the transformer blocks in the CTformer modules take Performer to facilitate model training. The embedding dimension for all transformer blocks is $64$. 
    \item In tokenization/detokenization, the kernel for the unfolding/folding is set to $7$ with a stride of $2$ to reduce computational cost. For the four CTformer modules, the cyclic shift strides in the T2TD/IT2TD blocks are $\{2, 2, -2, -2\}$. The kernel sizes of the unfolding/folding operations are 3 with dilations of $\{2, 1, 1, 2\}$, respectively. The strides are set to 1 to avoid the information loss. Thus, according to Eq. (\ref{unfold}), the corresponding token numbers $n_1$, $n_2$, and $n_3$ for the CTformer module A, CTformer module B, and the intermediate transformer layer are computed as follows:
\begin{equation}
    \begin{cases}
    n_1=  \Big ( \left \lfloor \frac{64-1\times (7-1)-1}{2}+1 \right \rfloor \Big ) ^2 = 841\\
    n_2=  \Big ( \left \lfloor \frac{\sqrt{841}-2\times (3-1)-1}{1}+1 \right \rfloor \Big ) ^2 = 625\\
    n_3=  \Big ( \left \lfloor \frac{\sqrt{625}-1\times (3-1)-1}{1}+1 \right \rfloor \Big ) ^2 = 529,
    \end{cases}
\end{equation}
here $\mathrm{spatial}(d) = \sqrt{841} = 29$ and $\mathrm{spatial}(d) = \sqrt{625} = 25$ can be calculated from the reshaping process in Eq. (\ref{reshape}). 
The transformer token numbers in the decoder are symmetrically arranged as $\{625, 841\}$.
%     \item In the encoder, the stride is $2$ for the cyclic shift layers. The kernel sizes for the unfolding operations (one from tokenization and two from CTformer modules)
%     % unfolding/folding operations in dilated unfolding stages 
%     are $\{7\times7$, $3\times3$, $3\times3$\} with strides of $\{2, 1, 1\}$ and dilations of $\{1, 2, 1\}$, respectively. The strides need to be small to prevent information loss from previous feature maps. However, if we set all the strides as $1$, the computational cost would be too high for model training. Therefore, only the first stride is $2$, while others are $1$. Thus, according to Eq. (\ref{unfold}), the corresponding token numbers $n_1, n_2$, and $n_3$ for the first three transformer blocks are computed as follows:
% \begin{equation}
%     \begin{cases}
%     n_1=  \Big ( \left \lfloor \frac{64-1\times (7-1)-1}{2}+1 \right \rfloor \Big ) ^2 = 841\\
%     n_2=  \Big ( \left \lfloor \frac{\sqrt{841}-2\times (3-1)-1}{1}+1 \right \rfloor \Big ) ^2 = 625\\
%     n_3=  \Big ( \left \lfloor \frac{\sqrt{625}-1\times (3-1)-1}{1}+1 \right \rfloor \Big ) ^2 = 529,
%     \end{cases}
% \end{equation}
% here $\mathrm{spatial}(d) = \sqrt{841} = 29$ and $\mathrm{spatial}(d) = \sqrt{625} = 25$ can be calculated from the reshaping process in Eq. (\ref{reshape}). 
% The transformer token numbers in the decoder are symmetrically arranged as $\{625, 841\}$.
\item We randomly extract $4$ patches from all available slices for training through $4000$ epochs with a batch size of $16$. In a training batch, fewer patches with more images lead to less fluctuations and bias than more patches with fewer images because many patches from a single image usually cannot represent the overall data distribution.

\item Adam is adopted to minimize the MSE loss with an initial learning rate of ${1.0\times 10^{-5}}$, which gradually decreases to ${1.0 \times 10^{-6}}$ with a scheduled decay rate. 
\item A margin size of $16$ is used for overlapped inference.
\end{itemize}

\textbf{Denoising performance.} The performance of the CTformer is compared to other state-of-the-arts, \textit{e.g.}, RED-CNN \cite{RN69}, WGAN-VGG \cite{RN68}, MAP-NN \cite{RN82}, and AD-NET \cite{RN55}. The selected models are all popular low-dose CT or natural image denoising models that were published in flagship journals. We retrain all the models based on their officially-disclosed codes. 

\begin{figure}
\centering
\includegraphics[width=\linewidth]{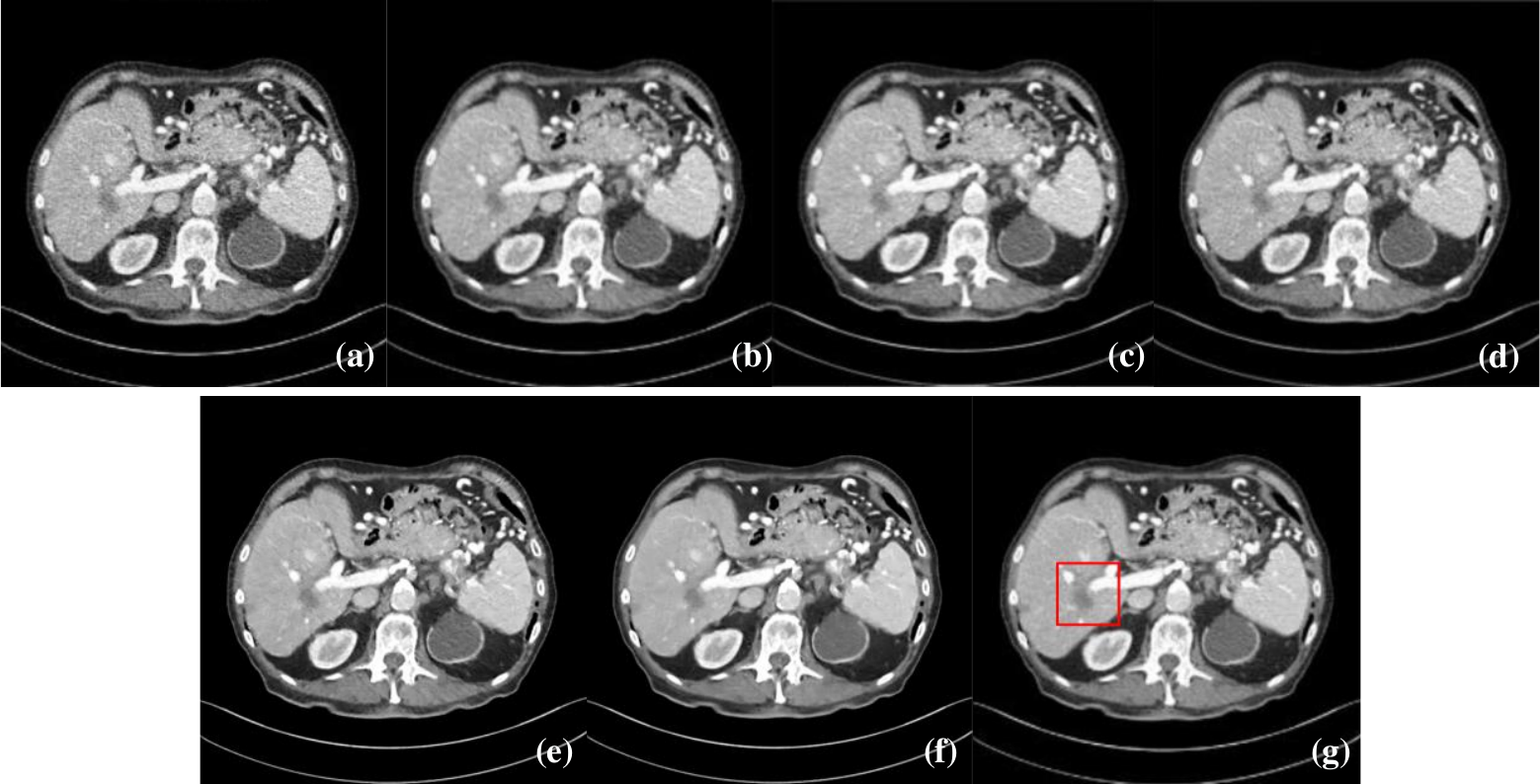}
\caption{The denoised results of different networks on L$506$ with Lesion No. $575$. (a) LDCT, (b) RED-CNN, (c) WGAN-VGG, (d) MAP-NN, (e) AD-NET, (f) the proposed CTformer, and (g) NDCT. The display window is [-160, 240] HU.} \label{fig:whole84}
\end{figure}

\begin{figure}
\centering
\includegraphics[width=\linewidth]{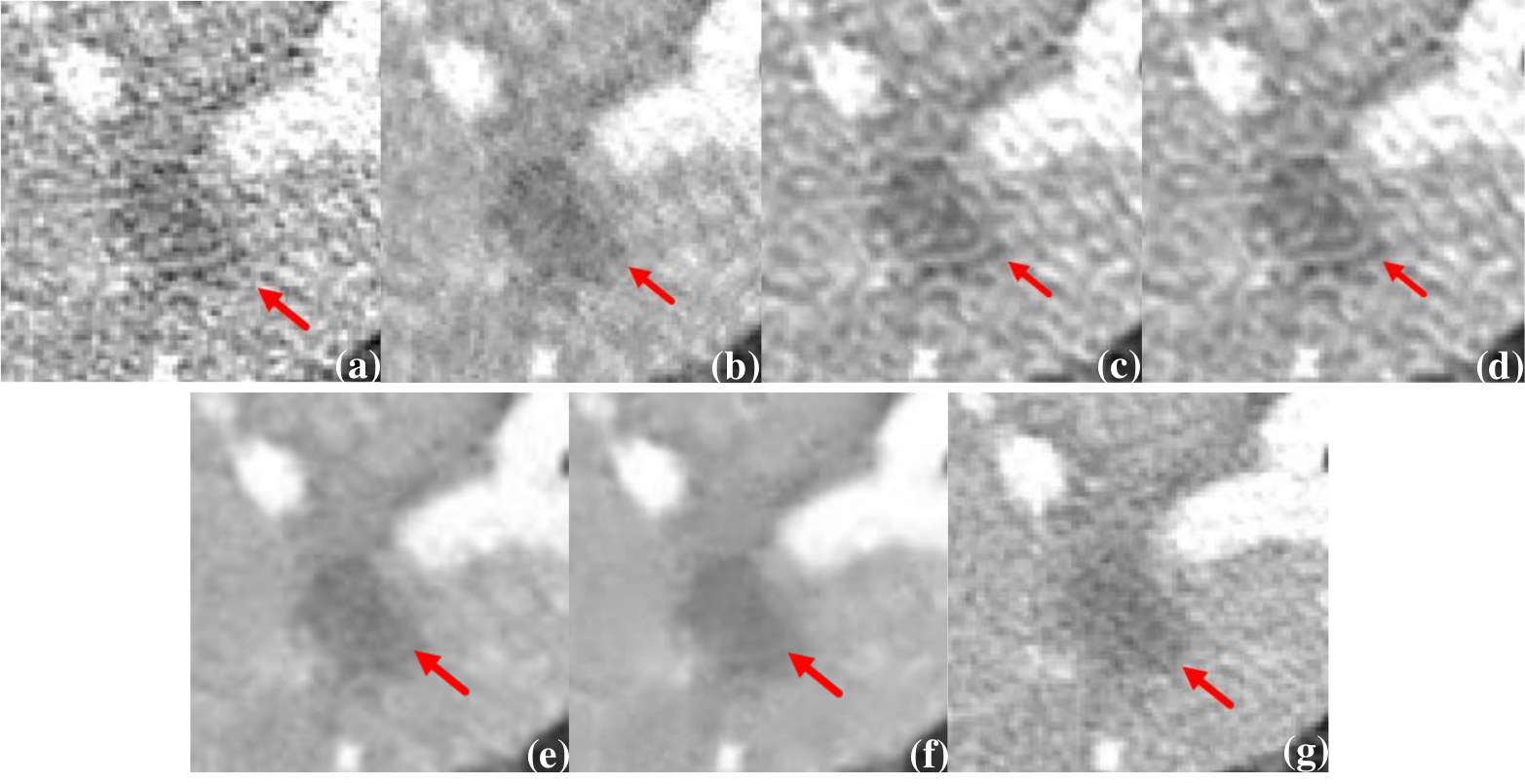}
\caption{The ROIs of the rectangle marked in Fig. \ref{fig:whole84}. (a) LDCT, (b) RED-CNN, (c) WGAN-VGG, (d) MAP-NN, (e) AD-NET, (f) the proposed CTformer, and (g) NDCT.} \label{fig:region84}
\end{figure}

Fig. \ref{fig:whole84} shows the results of different networks on L$506$ with Lesion No. $575$, and Fig. \ref{fig:region84} demonstrates the ROIs from the rectangular area marked in Fig. \ref{fig:whole84}. 
It can be seen that all methods can alleviate noise and artifacts to some extent, but the CTformer generates the clearest and the most perceptually-pleasing denoised images. Specifically, per the ROIs from Fig. \ref{fig:region84}, we find that WGAN-VGG and MAP-NN seem to introduce additional shadows and tissues. While the RED-CNN and AD-NET produce a smoother and clearer image relative to WGAN-VGG and MAP-NN, there still exists blotchy noise around the lesion. In contrast, the CTformer satisfactorily supresses the noise and artifacts, maintains high-level spatial smoothness, and keeps the structural details in the restored image. Therefore, we conclude that the CTformer is the best denoiser compared to its competitors. 

\begin{table}[h]
\caption{Quantitative evaluation results of different methods on L$506$ using SSIM and RMSE. The bold-faced numbers are the best results. }%\label{tab:all}
    %\begin{subtable}[h]{0.4\textwidth}
    \centering
    \begin{tabular}{l|c|c|c|c}%{p{0.2\textwidth}p{0.2\textwidth}p{0.2\textwidth}}
    \hline
    \hline
    Method &   \#param.  &  MACs &    SSIM$\uparrow$   &   RMSE$\downarrow$   \\
    \hline
    LDCT     & - & - &  0.8759 & 14.2416\\
    RED-CNN  & 1.85M & 5.05G &  0.9077 & 10.1044\\
    WGAN-VGG & 34.07M & 3.61G &  0.9008 & 11.6370\\
    MAP-NN   & 3.49M & 13.79G &  0.9084 & 9.2959\\
    AD-NET   & 2.07M & 9.49G &  0.9105 &  9.0997\\
    \hline
    CTformer & \textbf{1.45M}   &  \textbf{0.86G} &  \textbf{0.9121} &  \textbf{9.0233}\\  %9.0276
    \hline
    \end{tabular}
   \label{tab:quant}
    %\end{subtable}
    \hfill
     %\label{}
\end{table}
% \hspace{0.2cm}

% RED-CNN  & 1.85M & 5.05G &  0.8952 & 11.5926\\
% WGAN-VGG & 34.07M & 3.61G &  0.9008 & 11.6370\\
% MAP-NN   & 3.49M & 13.79G &  0.8941 & 11.5848\\
% AD-NET   & 2.07M & 9.49G &  0.9041 &  9.7166\\

Additionally, two metrics: \textit{structural similarity} (SSIM) and \textit{root mean square error} (RMSE) are adopted to quantitatively assess the quality of the denoised images. For fairness, we evaluate the model complexity with the number of trainable parameters (\#param.) and MACs. 
Table \ref{tab:quant} shows the average SSIM and RMSE on all slices of L506. Among the state-of-the-art methods, only AD-NET achieve an SSIM score over $0.91$, and only MAP-NN and AD-NET have an RMSE score below $10$. In contrast, our CTformer has the highest SSIM of $0.9121$ and smallest RMSE of $9.0233$.
Concerning model complexity, MAP-NN has the highest MACs of $13.79$G because it uses a lot of repeated modules, while WGAN-VGG has the greatest number of trainable parameters of $34.07$M because it uses VGG as a feature extractor. In contrast, the CTformer has the smallest number of parameters and the lowest MACs. 
Compared to its competitors, our model has the best performance with the lowest computational cost. 

\textbf{Model efficiency.}
Model efficiency is an important issue in deep learning. To further verify the model efficiency of the CTformer, we compare the CTformer with RED-CNN, MAP-NN and AD-NET by checking the SSIM and RMSE scores from different model sizes. For the CTformer, we change the model size by revising the embedding size of the intermediate transformer block. The embedding sizes are set to $\{64, 256, 512, 1024\}$, respectively. While for other models, we vary their sizes by using different number of filters in each layer. The filter numbers in RED-CNN, MAP-NN, and AD-Net are $\{64, 96, 128, 256\}$, $\{64, 128, 256, 400\}$, and $\{64, 96, 128, 256\}$, respectively.

\begin{figure}[h]
\centering
\includegraphics[width=\linewidth]{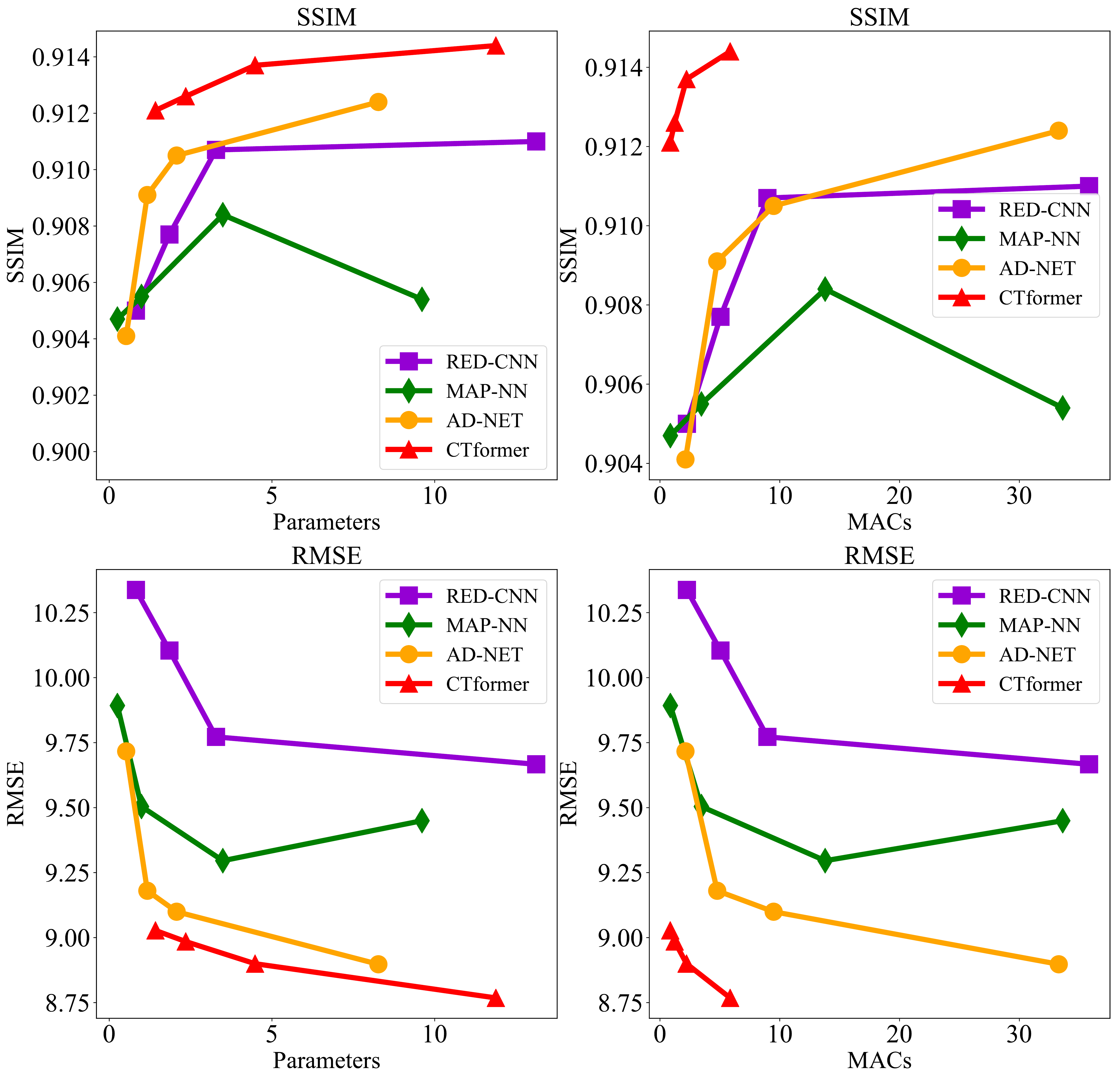}
\caption{The SSIM and RMSE curves of the CTformer and its competitors with respect to the number of parameters and MACs.} \label{model_size}
\end{figure}

Fig. \ref{model_size} shows the SSIM and RMSE scores of different models with respect to the number of parameters and MACs. The highlights of Fig. \ref{model_size} are that the SSIM curves of the CTformer lie on the top left of other curves, while its RMSE curves lies on the bottom left. When the number of parameters and MACs are close, the CTformer always delivers the best scores compared to the RED-CNN, MAP-NN and AD-NET. We summarize that the CTformer has the superior model efficiency to its competitors.

% \begin{figure}
% \centering
% \includegraphics[width=\linewidth]{figure/tsbl3.pdf}
% \caption{LOSS, SSIM and RMSE curves of CTformer-T, CTformer-S, CTformer-B, CTformer-L on case L$506$.}\label{fig:acc_curve_tsbl}
% \end{figure}

\begin{figure}
\centering
\includegraphics[width=\linewidth]{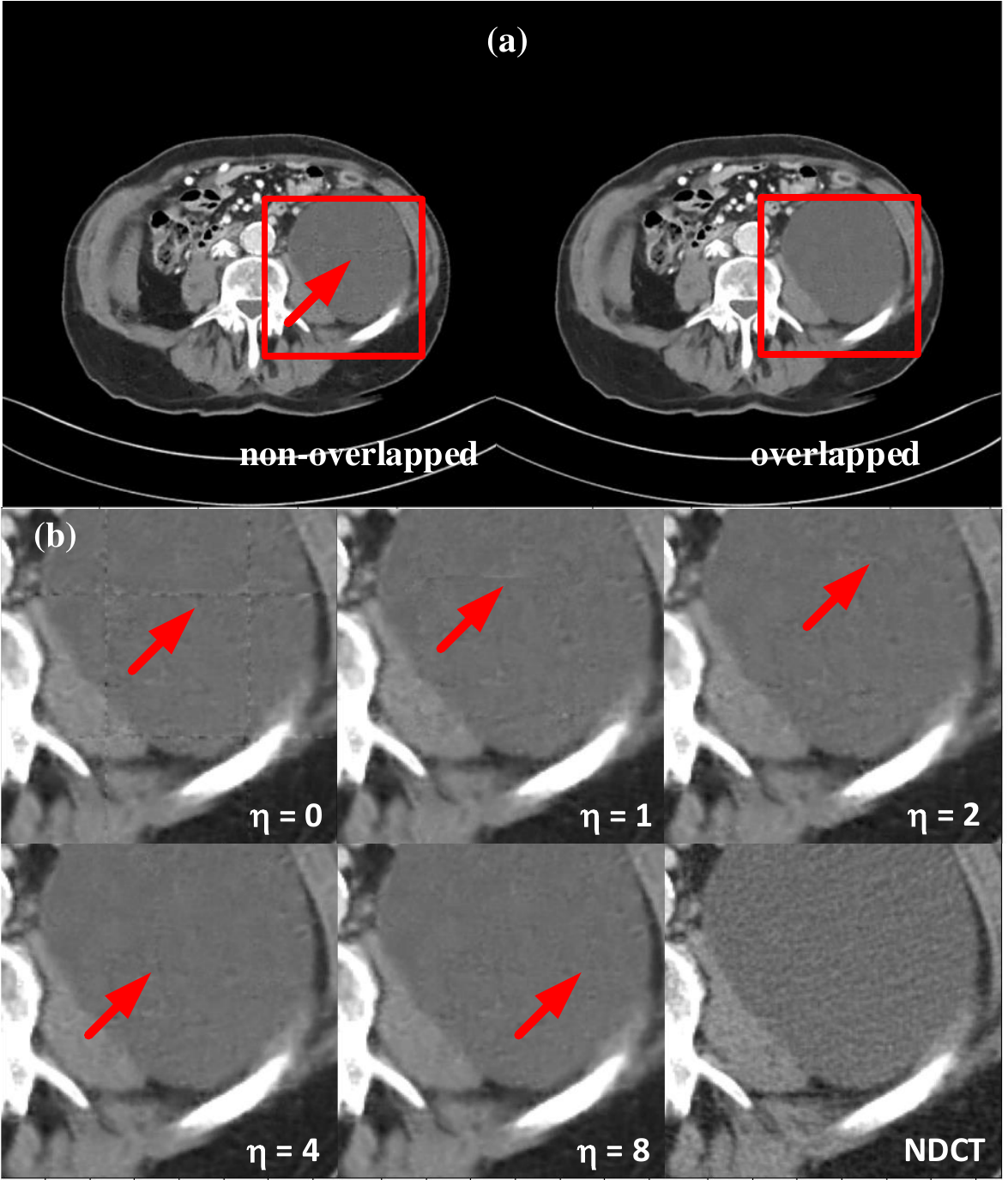}
\caption{(a) The denoised results of non-overlapped inference and overlapped inference. (b) The denoising results of different margin sizes on the ROIs indicated in (a).} \label{margin_r}
\end{figure}

\textbf{Eliminating boundary artifacts.}
The overlapped inference is performed to eliminate boundary artifacts as shown in Fig. \ref{margin_r}(a). From the ROIs in Fig. \ref{margin_r}(b), we can see that the boundary artifacts are obvious when $\eta$ is 0 or 1 but soon become hardly perceivable when $\eta$ further increases. It is worth noting that as $\eta$ varies, the boundary artifacts can appear in different regions because the size of the patches integrated in the final image is different.
To further confirm the effectiveness of the overlapped inference, quantitative analysis on the patient L$506$ is also conducted. As seen from Table \ref{margin_effect}, the SSIM and RMSE scores improve fast when $\eta$ goes from 0 to 20 with a better performance on $16$. The corresponding ratio of the extra computation over the authentic computation $\sigma$ is calculated from Eq.(\ref{ratio}): $\big( \frac{\lceil 512/(64-2 \times 16) \rceil}{\lceil 512/64 \rceil} \big )^2 = 4$. To sum up, the overlapped inference can sufficiently address the dense boundary artifacts. 

% \begin{figure}
% \centering
% \includegraphics[width=0.48\textwidth]{figure/overlap_all_v4.pdf}
% \caption{(a)The denoising results of different margin sizes on the ROI indicated by the circle in Fig. \ref{margin_r}(a). The red arrows indicate boundary artifacts.} \label{overlap_all}
% \end{figure}

% %previous most useful curve
% \begin{figure}
% \centering
% \includegraphics[width=0.95\linewidth]{figure/overlap_curve3.pdf}
% \caption{The SSIM/RMSE curves on patient L$506$ concerning different margin sizes $\eta$ for testing.} \label{overlap_quant}
% \end{figure}

\begin{table}[h]%{0.45\textwidth}
\centering
\caption{The SSIM and RMSE scores improve with margin.}
    \begin{tabular}{l|c|c|c|c|c|c}%{p{0.2\textwidth}p{0.2\textwidth}p{0.2\textwidth}}
        \hline
        \hline
        %\hspace{0.5cm}
        Margin  &    0   & 4 & 8 & 12 & 16 & 20    \\  
        \hline
        SSIM$\uparrow$  & 0.9071 & 0.9098 & 0.9113 & 0.9116 & \textbf{0.9121} & 0.9120 \\
        \hline
        % 0.9071,0.9087,0.9090,0.9093,0.9096,0.9098,0.9101,0.9108,0.9113,0.9113,0.9115,0.9115,0.9116,0.9115,0.9118,0.9119,0.9121,0.9121,0.9121,0.9120,0.9120   \\
        RMSE$\downarrow$ &  9.5671 & 9.1940 & 9.0890 & 9.0503 & 9.0233 & \textbf{9.0223} \\
        %9.5671,9.3292,9.2680,9.2312,9.1940,9.1651,9.1476,9.1137,9.0890,9.0804,9.0652,9.0548,9.0503,9.0462,9.0360,9.0318,9.0271,9.0233,9.0238,9.0217,9.0223
        \hline
    \end{tabular}
    \label{margin_effect}
\end{table}

\begin{figure*}
\centering
\includegraphics[width=0.9\linewidth]{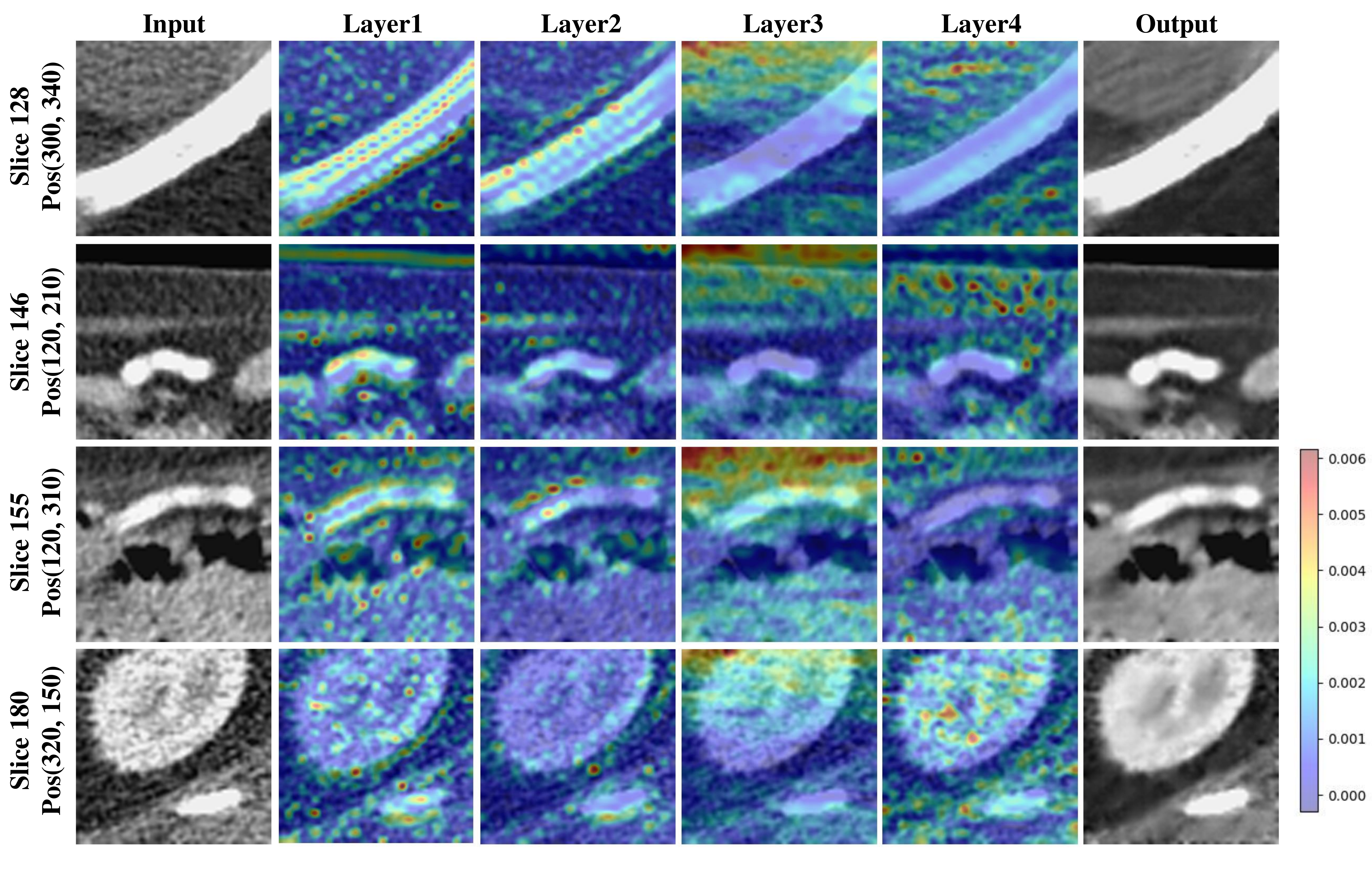}
\caption{The attention maps over different input slices on specific positions.}\label{attention_all}
\end{figure*}

\begin{figure*}
\centering
\includegraphics[width=\linewidth]{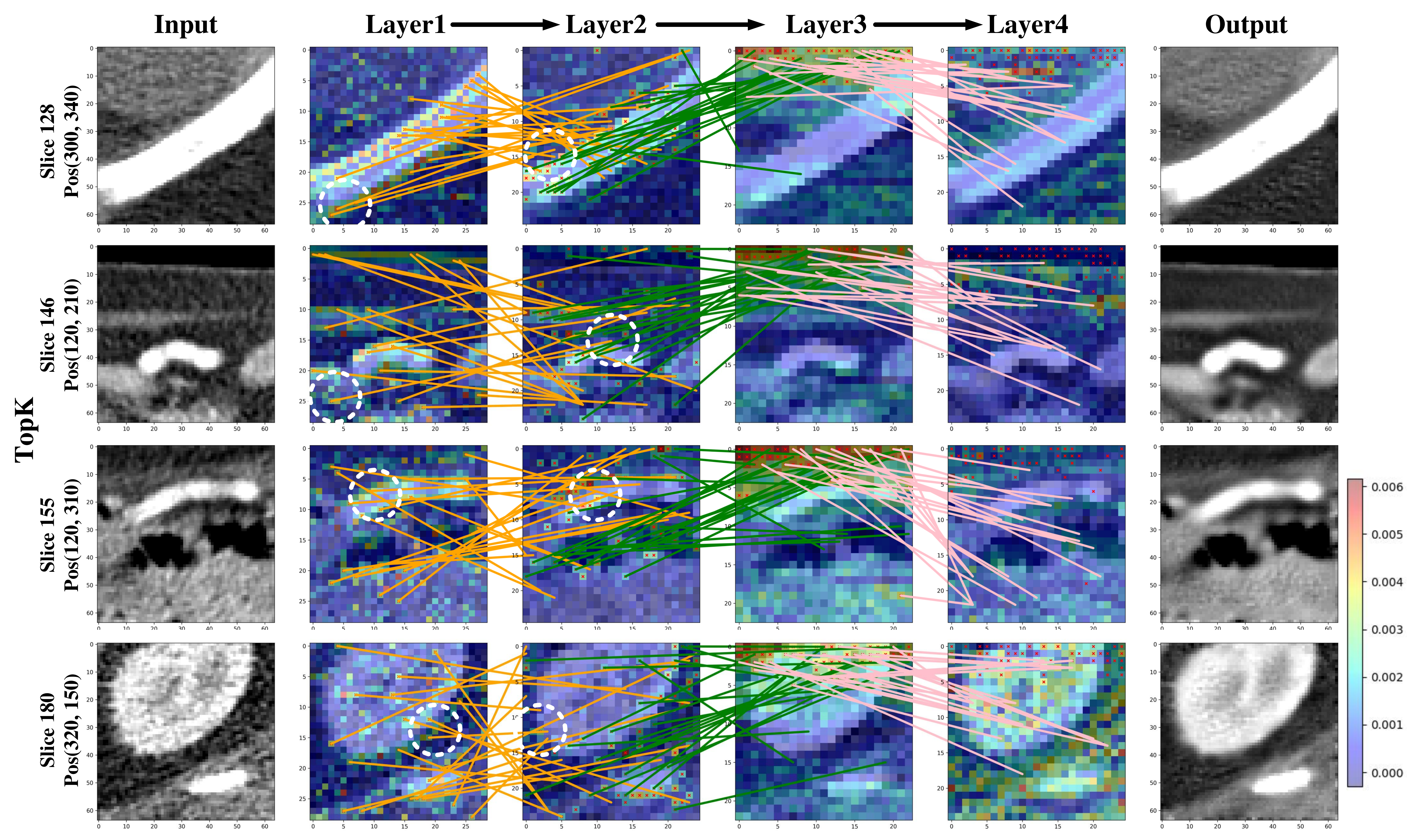}
\caption{TopK method for extraction of high activations and the corresponding attention graph.} \label{graph_attention}
\end{figure*}

\begin{figure*}
\centering
\includegraphics[width=\linewidth]{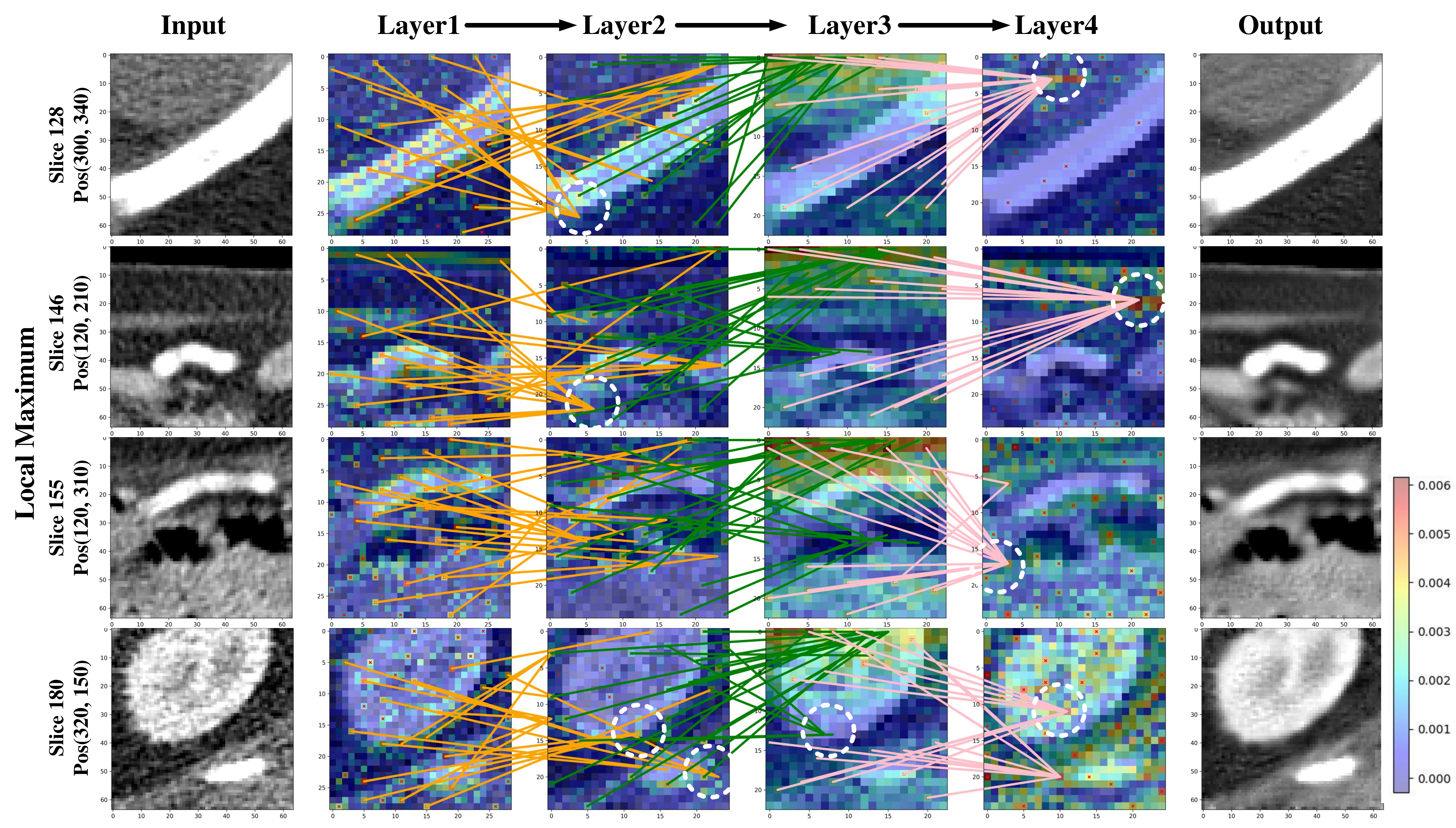}
\caption{Local maximum method for extraction of peak activations and the corresponding attention graph.} \label{graph_attention_LM}
\end{figure*}

\textbf{Visual interpretation.} To reveal the latent learning behavior in the CTformer, we visualize the attention maps $\mathrm{Att}=\mathrm{softmax}(\mathbf{QK}^\top / \sqrt{d_k})$ in each layer. Specifically, we derive attention maps by averaging all grids of $\mathrm{Att}$ and resize it to the size of the original image. Then, the attention map is superimposed on the image with a transparence rate 0.4. 

As shown in Fig. \ref{attention_all}, the attention map in the first layer highlights the key object parts. Specifically, there are more attentions on the edges rather than the composition of key structures like bones. Moreover, there are scattered dotted attentions on the protruding texture in the original image. For the attention map in the second layer, it basically resembles the pattern in the first attention, but sparser and less focused on the structures. Next, the pattern in the third layer becomes semantically implicit. Finally, the attention in the fourth layer tends to ignore the edges of objects and emphasize the content where noise is concentrated. 

% As indicated by the purple circle in Fig. \ref{attention_all}, the first and fourth attention maps highlight most key structures including the marked lesions.

% For the last attention, as pointed out by the bidirectional arrow in Lesion.54's 'Attention5' image, we can see the boundary artifacts which we will address using overlapped inference. Moreover, we can also obtain the multiple head attention (MHA) layer by multiplying the 'Attention5' matrix with the corresponding \textit{Value5}. By investigating the MHA map, we can find that it can also more or less learn the noise pattern. As shown by the white circle in Fig. \ref{attention_all}, the unique structures of the noise can be recognized from the MHA layer. In summary, the attention in the CTformer can effectively discriminate the key structures and the noise pattern in an interpretable and transparent way. 

Since attentions in different layers focus on different structures, we construct an explanatory graph to illustrate the flow of attention across various layers. In our experiments, the object nodes are represented by the pixel coordinates of the image. We select the top 60 activations in the attention maps as nodes using TopK/LM selection and identify the highest activation under each node's influence. By applying the proposed method, the whole TopK and LM graph are obtained in Figs. \ref{graph_attention} and \ref{graph_attention_LM}, respectively. 

From the TopK graph in Fig. \ref{graph_attention}, it can be seen that the attention flows across different testing slices have very similar patterns. First, from the first to the second layer, the attentions on the edges still favor other edges in the next layer as indicated by the white circle in Fig. \ref{graph_attention}. Second, all high activations from the second layer move to the top area of the third layer in a latent manner. Last, all the top attentions in the third layer spread across the noisy area in the fourth layer. While the TopK graph identifies the flow of the top activations, the LM graph illustrates that of the local protuberant objects. As shown in Fig. \ref{graph_attention_LM}, the attention graphs of different slices using LM are also analogous. Compared to TopK graph, one principal distinction in the LM graph is that groups of local maximum activations tend to implicitly concentrate on the same point in the next layer. The white circles in Fig. \ref{graph_attention_LM} illustrate some concurrent points. Therefore, by inspecting the two attention graphs, the dynamic flow can be clearly followed. We can figure out how the object parts are co-activated and thus go through similar level of noise reduction.

In summary, the latent learning behavior of the CTformer can be visually interpreted statically and dynamically. This makes the proposed model more transparent and reliable for diagnostic decisions.
% while Fig. \ref{graph_attention_LM} shows the graph with LM for part of the graph from which the attention flow between the third and the fourth attention can be traced. In this way, the latent representations of the attentions in the CTformer are revealed. This makes our model more transparent in making the diagnostic decisions.

\section{ABLATION STUDY}

In this part, comparative experiments are conducted to study the impact of the T2TD block, the cyclic shift operation and the number of the intermediate transformer blocks.  
%\textbf{\textit{On the influence of cyclic shift:}}
\begin{figure*}
\centering
\includegraphics[width=0.8\textwidth]{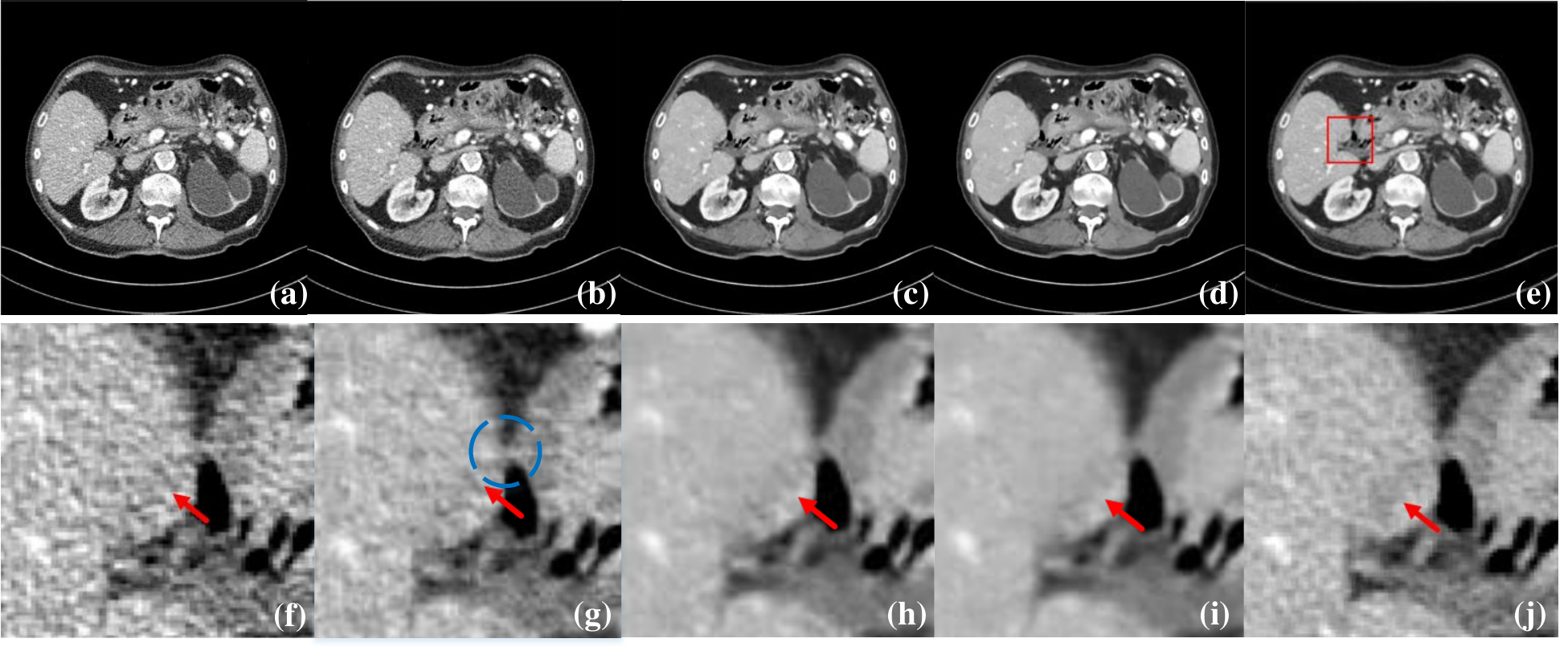} %v3
\caption{The performance of CTformer on case L$506$ with lesion No. $576$. (a) LDCT, (b) Solve-ViT (c) CTformer without cyclic shift, (d) CTformer, and (e) NDCT. (f)-(j) are the corresponding magnified ROIs from (a)-(e).}
\label{ablation}
\end{figure*}

\begin{figure}
\centering
\includegraphics[width=\linewidth]{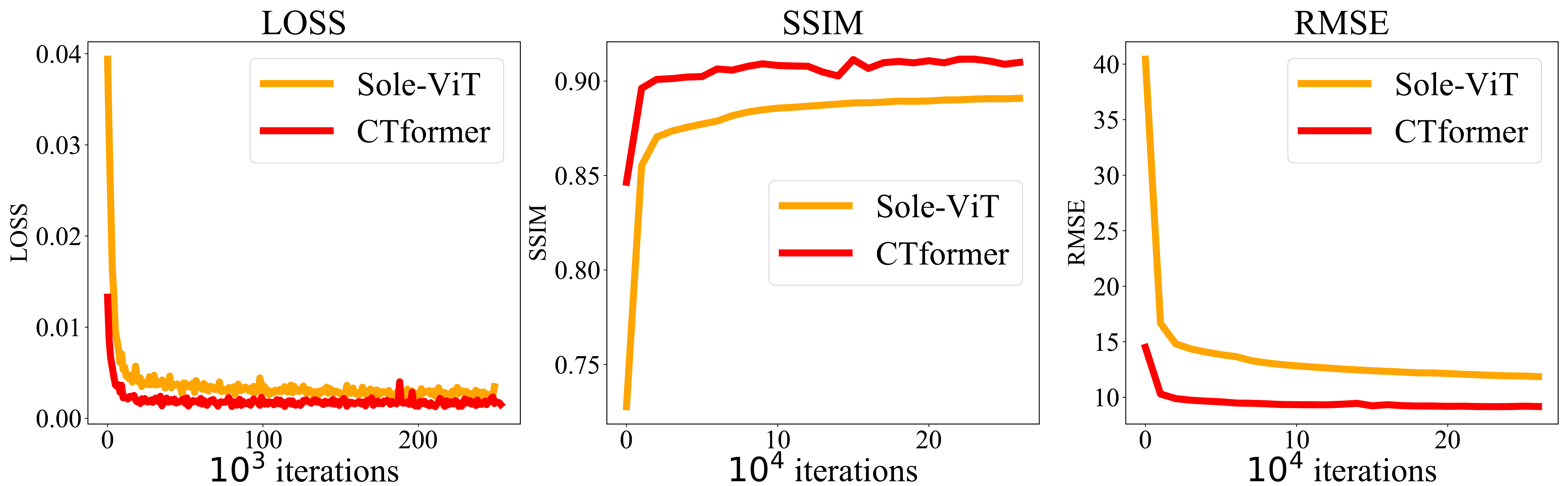}
\caption{Visualization of LOSS, SSIM and RMSE curves of CTformer and Sole-ViT on case L$506$ after different iterations.}\label{acc_curve-vit}
\end{figure}

\textbf{Impact of T2TD block.}
T2TD blocks are used in the CTformer to enhance the feature integration in the tokenization stage. Compared to fixed-region tokenization, the tokens in T2TD blocks are extracted from various regions of the original images. 
% If we do not use T2T modules in the model, the tokens fed into the transformers are tokenized from a fixed region of the image. Theoretically speaking, our model is more powerful in combining features from different subareas of the input images. 
To verify the effectiveness of this part, a Sole-ViT model without the T2TD module is designed. We only adopt a sole convolution in the tokenization stage with a filter size of $8$ and a stride of $8$. Then five layers of transformer with an embedding size of $256$ are applied for feature extraction and denoising. $256$ rather than $64$ embedding size is used because the model size and MACs are close to our model as shown in \ref{tab:quantTB}. 
% Table \ref{tab:quant2}.
% with a little bit more parameters and less MACs as shown in Table \ref{tab:quant2}.
Finally, a detokenization with deconvolution is employed to transform the tokens back to desired image domain. By investigating the conjunction area inside the blue circle in Fig. \ref{ablation}(g), we can see that Sole-ViT brings in extra blotchy tissues. Meanwhile, Fig. \ref{acc_curve-vit} shows that the CTformer converges faster than Sole-ViT and has better scores with a margin of $0.0235$ on SSIM and $3.3362$ on RMSE. 

\textbf{Impact of cyclic shift.}
In this work, the cyclic shift is performed in the T2TD blocks to enhance the perceptual fields of our model. Fig. \ref{ablation} shows that CTformer with cyclic shift enjoys more spatial smoothness compared to the CTformer without cyclic shift. The latter introduces some additional noise components. Quantitative results from Table \ref{tab:quantTB} also confirm the effectiveness of cyclic shift in improving the SSIM and RMSE of the model by $0.0026$ and $0.1337$, respectively. 

\textbf{Impact of block number.}
In terms of the number of intermediate transformer blocks, we evaluate the CTformer with $1$, $2$, $4$, and $8$ blocks to identify the influence. When the block number grows, the network goes deeper. The computational cost increases slowly, but the actual training time climb up dramatically. However, Table \ref{tab:quantTB} indicates that the CTformer with only one block yields the best performance over the ones with more blocks. 
% This demonstrates that shallow network suites more for the LDCT denoising task than the deep network. 

\begin{table}[h]%{0.45\textwidth}
\centering
\caption{Quantitative evaluation results of the Sole-ViT, the CTformer(W/oCS), and the CTformers with different number of transformer blocks.}
    \begin{tabular}{|l|c|c|c|c|c|}%{p{0.2\textwidth}p{0.2\textwidth}p{0.2\textwidth}}
        \hline
        %\hspace{0.5cm}
        Method & TB  &    \#param.   &   MACs  &    SSIM$\uparrow$   &   RMSE$\downarrow$   \\
        \hline
        Sole-ViT  &  1 & 2.92M  &  \textbf{0.24G}  & 0.8886 & 12.3595\\
        CTformer(W/oCS) & 1 & \textbf{1.45M}   &  0.86G & 0.9095 & 9.1570\\
        \hline
        CTformer & 1   & \textbf{1.45M}  &  0.86G  & \textbf{0.9121} & \textbf{9.0233}\\
        CTformer & 2   & 1.48M  &  0.87G  & 0.9115 & 9.0303\\
        CTformer & 4   & 1.55M  &  0.91G  & 0.9108 & 9.1285\\
        CTformer & 8   & 1.68M  &  0.98G  & 0.9115 & 9.0841\\
        \hline
        
    \end{tabular}
    \label{tab:quantTB}
\end{table}

\section{CONCLUSION}
In this paper, we have proposed a novel convolution-free transformer empowered by dilated tokenization and cyclic shift for LDCT denoising, which is referred to as the CTformer. To the best of our knowledge, the proposed CTformer is the first pure transformer model for LDCT denoising. Also, we have developed the interpretation methods for the proposed CTformer to decode its hidden behavior. Moreover,
we have proposed the overlapped inference to address the boundary artifacts that are common in an encoder-decoder model. Experimental results have demonstrated that the CTformer outperforms its competitors in terms of the denoising performance and model efficiency. In the future, more efforts can be made to translate the CTformer into other medical denoising problems.

\bibliographystyle{ieeetr}
\bibliography{CTformer.bib}
\end{document}